\documentclass[useAMS,usenatbib,usegraphicx]{mn2e}

\usepackage{amssymb}

\newcommand{\etal}{{et al.~}}
\newcommand{\kms}{\>{\rm km}\,{\rm s}^{-1}}

\title{GMOS IFU observations of the stellar and gaseous kinematics
in the centre of NGC~1068}

\author[J. Gerssen et al.]
{Joris Gerssen,$^{1}$\thanks{E-mail:
joris.gerssen@durham.ac.uk}
Jeremy Allington-Smith,$^{1}$
Bryan W. Miller,$^{2}$
\newauthor
James E. H. Turner$^{2}$
and Andrew Walker$^{1}$
\thanks{Current address: e2v technologies (uk) ltd, 106
Waterhouse Lane, Chelmsford, Essex, CM1 2QU, UK}\\
$^{1}$University of Durham, Department of Physics, South Road, Durham
       DH1 3LE, UK \\
$^{2}$Gemini Observatory, La Serena, Chile}

\begin{document}

\date{Accepted Received}

\pagerange{\pageref{firstpage}--\pageref{lastpage}} \pubyear{2005}

\maketitle

\label{firstpage}

\begin{abstract}

We present a datacube covering the central 10 arcsec of the archetypal
active galaxy NGC~1068 over a wavelength range 4200 -- 5400 \AA \
obtained during the commissioning of the integral field unit (IFU) of
the Gemini Multiobject Spectrograph (GMOS) installed on the
Gemini-North telescope.  The datacube shows a complex emission line
morphology in the [O{\small III}] doublet and H$\beta$ line.  To
describe this structure phenomenologically we construct an atlas of
velocity components derived from multiple Gaussian component fits to
the emission lines. The atlas contains many features which cannot be
readily associated with distinct physical structures.  While some
components are likely to be associated with the expected biconical
outflow, others are suggestive of high velocity flows or disk-like
structures. As a first step towards interpretation, we seek to
identify the stellar disk using kinematical maps derived from the
Mg\,$b$ absorption line feature at 5170 \AA \ and make associations
between this and gaseous components in the atlas of emission line
components.
\end{abstract}


%%%%%%%%%%%%%%%%%%%%%%%%%%%%%%%%%%%%%%%%%%%%%%%%%%%%%%%%%%%%%%%%%%%%%%%%

\section{Introduction}

The galaxy NGC~1068 is the nearest example of a system with an active
galactic nucleus (AGN).  Integral field spectroscopy allows the
complex kinematics of this object to be revealed without the
ambiguities or errors inherent in slit spectroscopy (e.g. Bacon 2000)
and provides high observing efficiency since all the spectroscopic
data is obtained simultaneously in a few pointings. The use of fibre
bundles coupled to close-packed lenslet arrays offers not only high
throughput for each spatial sample, but also unit filling factor to
maximise the overall observing efficiency and reduce the effect of
sampling errors. The GMOS Integral Field Unit, a module which converts
the Gemini Multiobject Spectrograph (Allington-Smith \etal 2002)
installed at the Gemini-north telescope, from slit spectroscopy to
integral field spectroscopy (the first such device on a 8-10m
telescope) offers an opportunity to gather data on this archetypal
object at visible wavelengths in a completely homogeneous way with
well-understood and unambiguous sampling.

In this paper we present a datacube covering the central $10 \times
8$~arcsec of NGC~1068 over a wavelength range from 4200 -- 5400 \AA \
obtained during the commissioning run of the GMOS IFU in late 2001.

\begin{figure}
\includegraphics[width=8cm]{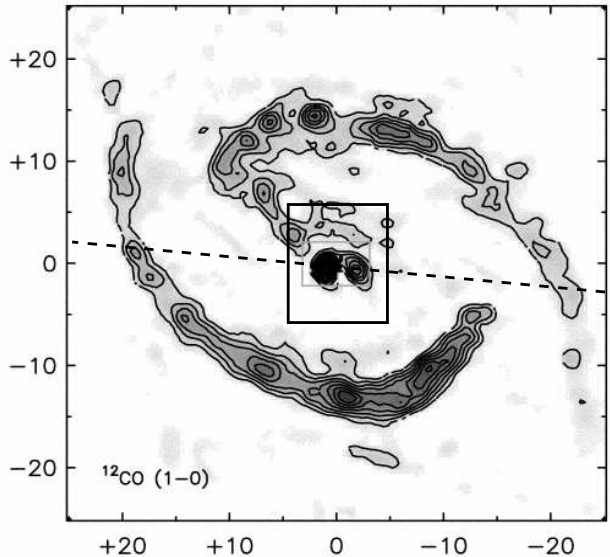}
\caption{
The field-of-view ($10.3 \times 7.9$ arcsec) of the mosaicked GMOS IFU
data is outlined (black rectangle) on a CO map (with the axes is
arcsec) of the centre of NGC~1068 reproduced from Schinnerer \etal
2000. The CO map illustrates the central complexity in this galaxy
with an inner spiral, an inner bar (PA$=48^\circ$) and a central CO
ring.  Also indicated is the position angle of the galaxy's major axis
(dashed line) at PA$=80^\circ$. At the assumed distance of 14.4 Mpc to
NGC~1068, 1 arcsec corresponds to 70 pc. }
\label{f:fov}
\end{figure}

Below we briefly summarize the key morphological characteristics of
NGC~1068 (see also Antonucci, 1993; Krolik \& Begelman,
1988). Although the Seyfert galaxy NGC~1068 resembles an ordinary Sb
type galaxy at its largest scales, the central region exhibits
considerable complexity.  Figure~\ref{f:fov} shows a CO map reproduced
from Schinnerer \etal (2000) detailing the structure on a $1$ kpc
scale.  Two inner spiral arms and the inner stellar bar first
recognized by Scoville \etal (1988, see also Thatte et al., 1997) can
clearly be seen.  The position angle of the galaxy's major axis
(measured at $D_{\rm 25} = 70$ arcsec) is indicated by the dashed
line.  The central CO ring with a diameter of about five arcsec is
roughly aligned with this axis.  The field covered by our GMOS-IFU
observations is indicated in the figure.  The galaxy's distance
throughout this paper is assumed to be 14.4 Mpc, yielding a scale of
70 pc per arcsec.

At subarcsecond scales, radio interferometer observations (Gallimore
\etal 1996) show a triple radio component with a NE approaching
synchrotron-emitting jet and a SW receding jet.  Hubble Space
Telescope observations (HST) by Macchetto \etal (1994) in [O{\small
III}] emission show a roughly North-South oriented v-shaped region
with various sub-components.  With the exception of the central
component, there seems to be no immediate correspondence between the
[O{\small III}] components and the radio components.  More recent HST
observations (Groves et al. 2004; Cecil et al. 2002) reveal the
presence of numerous compact knots with a range of kinematics and
inferred ionising processes.  Recent infrared interferometer
observations (Jaffe \etal 2004) have identified a 2.1$\times$3.4 pc
dust structure that is identified as the torus that hides the central
AGN from view (e.g. Peterson, 1997).

In section 2, we give details of the observations and construction of
the datacube. The datacube is analysed to decompose the emission lines
into multiple components in Section 3, which also includes a brief
examination of the main features that this reveals. In section 4, we
use the stellar absorption features to constrain a model of the
stellar disk and make associations between this and components in the
emission line analysis.

%%%%%%%%%%%%%%%%%%%%%%%%%%%%%%%%%%%%%%%%%%%%%%%%%%%%%%%%%%%%%%%%%%%%%%%%

\section{Observations and Data Reduction}

\begin{figure*}
\includegraphics[width=18cm]{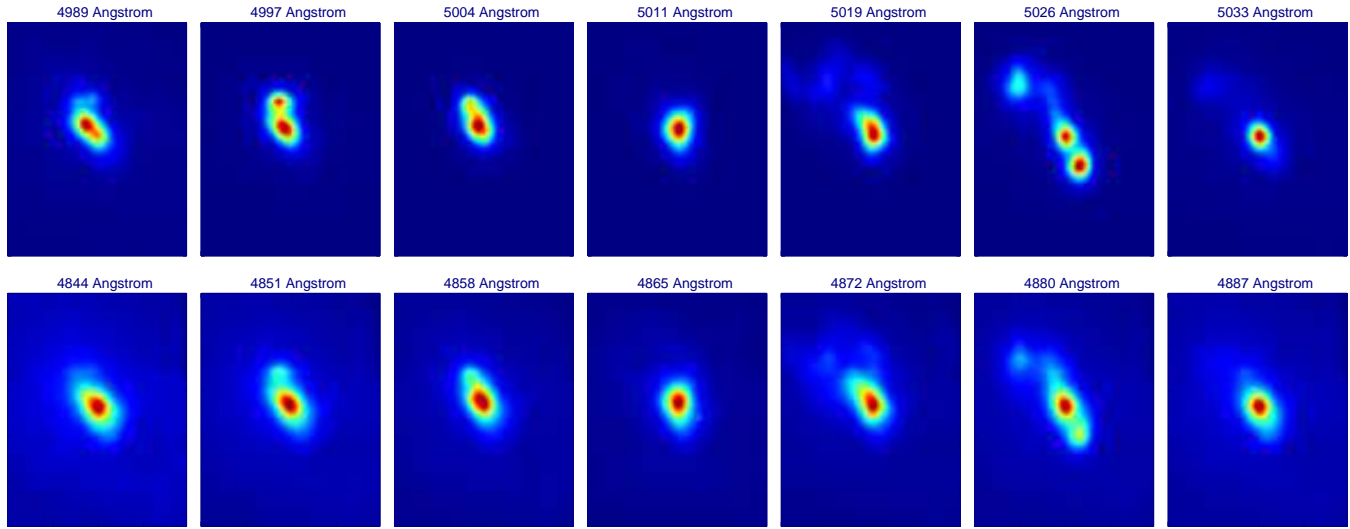}
\caption{The top panels (each covering an area of 10.3 x 7.3 arcsec)
show a series of one Angstrom wide wavelength slices through the data
across the brightest [O{\small III}] line.  Across this line the
central morphology of NGC~1068 develops various subcomponents.  A
similar change in morphology is witnessed across all other emission
lines as illustrated by the next series of panels that show the
emission distribution across the H$\beta$ line.}
\label{f:frames}
\end{figure*}

The GMOS IFU (Allington-Smith \etal 2002) is a fiber-lenslet system
covering a field-of-view of $5 \times 7$ arcsec with a spatial
sampling of 0.2 arcsec. The IFU records 1000 contiguous spectra
simultaneously and an additional 500 sky spectra in a secondary field
located at 1.0 arcmin from the primary field.

The NGC~1068 observations on 9 Sep 2001 consist of 4 exposures of
900s.  obtained with the B600 grating set to a central wavelength of
4900 \AA \ and the $g'$ filter (required to prevent overlaps in the
spectra from each of the two pseudoslits into which the field is
reformatted).  The wavelength range in each spectrum was 4200 -- 5400
\AA. The seeing was ~0.5 arcsec. Between exposures the telescope
pointing was offset by a few arcsec to increase the covered area on
the sky and to improve the spatial sampling. The pointing offsets from
the centre of the GMOS field followed the sequence: (-1.75,+1.50),
(-1.75,-3.00), (+3.50,+0.00), (+0.00,+3.00) arcsec.

The data were reduced using pre-release elements of the GMOS data
reduction package (Miller et al. 2002 contains some descriptions of
the IRAF scripts).  This consisted of the following stages.

\begin{itemize}

\item {\it Geometric rectification}. This joins the three individual
CCD exposure (each of $4608 \times 2048$ pixels) into a single
synthesised detector plane, applying rotations and offsets determined
from calibration observations. At the same time, a global sensitivity
correction is applied to allow for the different gain settings of the
CCDs. Before this, the electronic bias was removed.

\item {\it Sensitivity correction}.  Exposures using continuum
illumination provided by the Gemini GCAL calibration unit were
processed as above.  GCAL is designed to simulate the exit pupil of
the telescope in order to remove calibration errors due to the
illumination path.  Residual large-scale errors were very small and
were removed with the aid of twilight flatfield exposures.  The mean
spectral shape and the slit function (which includes the effect of
vignetting along the direction of the pseudo-slit) were determined and
used to generate a 2-D surface which was divided into the data to
generate a sensitivity correction frame which removes both pixel-pixel
variations in sensitivity, including those due to the different
efficiencies of the fibres, and longer-scale vignetting loss in the
spatial direction.

\item {\it Spectrum extraction}. Using the continuum calibration
exposures, the 1500 individual spectra were traced in dispersion
across the synthesised detector plane. The trace coefficients were
stored for application to the science exposures. Using these, the
spectra were extracted by summing with interpolation in the spatial
direction a number of pixels equal to the fibre pitch around the mean
coordinate in the spatial direction.  Although the spectra overlap at
roughly the level of the spatial FWHM, resulting in cross-talk between
adjacent fibres at the pseudoslit, the effect on spatial resolution is
negligible since fibres which are adjacent at the slit are also
adjacent in the field and the instrumental response of each spaxel is
roughly constant. Furthermore it has no impact on the conditions for
Nyquist sampling since this is determined at the IFU input (where
$\geq 2$ spaxels samples the seeing disk FWHM) and not at the
pseudo-slit. However the overlaps permit much more efficient
utilisation of the available detector pixels resulting in a larger
field of view for the same sampling (see Allington-Smith and Content
1998 for further details).

\item {\it Construction of the datacube}. Spectra from the different
pointings were assembled into individual datacubes. The spectra were
first resampled onto a common linear wavelength scale using a
dispersion relationship determined from 42-47 spectral features
detected in a wavelength calibration observation. From the fitting
residuals we estimate an RMS uncertainty in radial velocity of 8
km/s. The offsets between exposures were determined from the centroid
of the bright point-like nucleus in each datacube, after summing in
wavelength. The constituent datacubes were then co-added using
interpolation onto a common spatial grid. The spatial increment was
$0.1 \times 0.1$ arcsec to minimise sampling noise due to both the
hexagonal sampling pattern of the IFU and the offsets not being exact
multiples of the IFU sampling increments.  The resampling does not
significantly degrade spatial resolution since the resampling
increment is much smaller than the Nyquist limit of the IFU. Since the
spectra had a common linear wavelength scale within each datacube, no
resampling was required in the spectral domain.  Cosmic ray events
were removed by identifying pixels where the data value significantly
exceeded that estimated from a fit to neighbouring pixels.  The
parameters of the program were set conservatively to avoid false
detections.  Since each spectrum spans $\sim$5 pixels along the slit,
distinguishing single-pixel events such as cosmic rays from emission
lines which affect the full spatial extent of the spectrum is quite
simple.  After this the exposures were checked by eye and a few
obvious low-level events removed manually using local interpolation.

\item {\it Background subtraction}. Spectra from the offset field were
prepared in the same way as for the object field above, combined by
averaging and used to subtract off the background sky signal. The sky
lines were relatively weak in this observation so the accuracy of this
procedure was not critical.

\end{itemize}

The result is a single merged, sky-subtracted datacube where the value
at each point is proportional to the detected energy at a particular
wavelength, $\lambda$, for a $0.1\times 0.1$ arcsec sample of sky at
location $x,y$. The fully reduced and mosaicked NGC~1068 GMOS IFU data
cube covers an area on the sky of $10.3 \times 7.9$~arcsec with a
spatial sampling of 0.1 arcsec per pixel and spectral resolving power
of 2500. The wavelength range covered was 4170 -- 5420 \AA \ sampled
at 0.456 \AA \ intervals.

The change in atmospheric dispersion over the wavelength range studied
is less than 1 spaxel so no correction has been applied. The spectra were not
corrected to a scale of flux density since this was not required for
subsequent analysis

%%%%%%%%%%%%%%%%%%%%%%%%%%%%%%%%%%%%%%%%%%%%%%%%%%%%%%%%%%%%%%%%%%%%%%%%

\section{Emission Line Data}

The morphology of the central region changes rapidly across the
emission lines.  This is illustrated in Fig.~\ref{f:frames} showing a
series of monochromatic slices. The emission lines appear to consist of
multiple components whose relative flux, dispersion and radial velocity
change with position.

\subsection{Empirical multi-component fits}

To understand the velocity field of the line-emitting gas, we made
multicomponent fits to the H$\beta$ emission line. This was chosen in
preference to the brighter O[{\small III}] $\lambda\lambda$ 4959, 5007
doublet because of its relative isolation and because the signal/noise
was sufficient for detailed analysis.  The fits were performed on the
spectra extracted for each spatial sample from the sky-subtracted and
wavelength-calibrated datacube after it had been resampled to $0.2
\times 0.2$ arcsec covering a field of 
$8.6 \times 6$~arcsec and 4830 -- 4920 \AA \ in wavelength to
isolate the H$\beta$ line. The resampling made the analysis easier by 
reducing the number of datapoints without loss of spatial information.

The continuum was found to be adequately fit by a linear function
estimated from clean areas outside the line profile. A program was
written to fit up to 6 Gaussian components, each characterised by
its amplitude, $A$, radial velocity, $v$ and velocity dispersion,
$\sigma$. The {\it Downhill Simplex} method of Nelder and Mead (1965)
was
used. This minimises $\chi^2$ evaluated as
\begin{equation}
\chi^2 = \sum_{i=1}^M \left(
\frac{ F_i - f(x_i ; A_i , \mu_i ,\sigma_i ... A_N, \mu_N , \sigma_N) }{
s_i }
                        \right)
\end{equation}
where $F_i$ is the value of the $i$th datapoint. $M$ is the number of
datapoints in the fit and
\begin{equation}
f(x_i)   = \sum_{j=1}^N
A_j \exp \left[ - \frac{ (x_i-\mu_j)^2 }{ 2 \sigma_j^2 } \right]
\end{equation}
is the sum of $N$ Gaussian functions. The radial velocity and velocity
dispersion are determined from $\mu$ and $\sigma$ respectively.  The
noise, $s$, in data numbers was estimated empirically from the data as
a sum of fixed and photon noise as

\begin{equation}
s_i = H\sqrt {\frac{F_i}{G} + s_R^2}
\end{equation}
where the detector gain $G = 2.337$ and the readout noise is $s_R =
3.3$. The parameter $H=1.06$ represents the effect of unknown additional
sources of error and was chosen to provide satisfactory values of $Q$
(see below) for what were judged by eye to be good fits.

The significance of the fit was assessed via $Q(\chi^2 | \nu)$ which is
the probability that $\chi^2$ for a correct model will exceed the
measured $\chi^2$ by chance. The number of degrees of freedom, $\nu =
M-3N$.  $Q$ is evaluated as
\begin{equation}
Q(\chi^2 | \nu) = Q(\nu/2, \chi^2/2) =
\frac{1}{ \Gamma(\nu/2) } \int^\infty_{\chi^2/2} \exp (-t) t^{\frac{\nu}
{2}-1} dt
\end{equation}
with limiting values $Q(0 | \nu) = 1$ and
$Q(\infty| \nu) = 0$.

Fits were attempted at each point in the field with $1 \leq N \leq
N_{\rm max} = 6$, with $N$ increasing in unit steps until $Q \simeq
1.0$. The fits were additionally subject to various constraints to
preclude unphysical or inherently unreliable fits, such as those with
line width less than the instrumental dispersion ($\sigma < 0.9$ pixels)
or so large as to be confused with errors in the baseline subtraction
($\sigma > 0.5 M $ ). The final value of the fit significance (not
always) unity was recorded as $Q_{\rm max}$. The fits at each datapoint
were examined visually and in the small number of cases, where the fits
were not satisfactory, a better fit was obtained with an initial choice
of  parameters chosen by the operator.

\begin{figure}
\includegraphics[width=10cm]{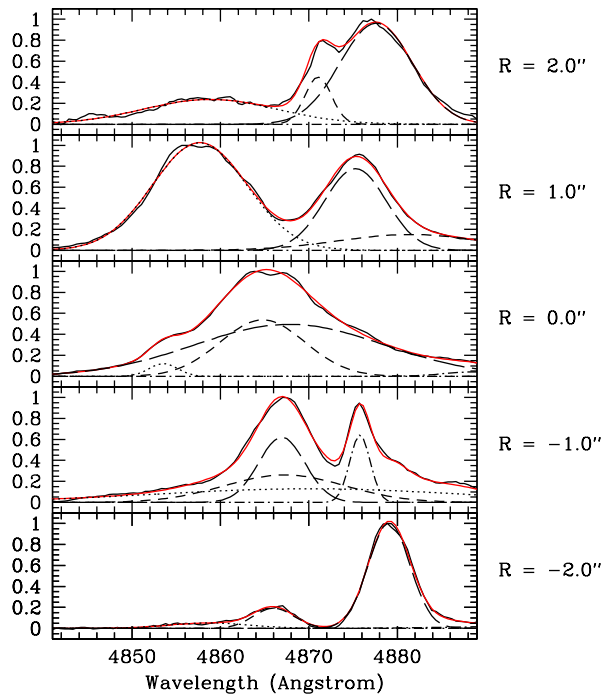}
\caption{The normalized emission line profiles (black) obtained at
various locations along a transect that coincides with the position
angle of the jets.  The individual Gaussian components whose sum
best-fits the observed line profile are shown in different line
styles. The red line shows the sum of the individual components.}
\label{f:transect}
\end{figure}

The uncertainty in the resulting radial velocities taking into account
random and systematic calibration errors was estimated to be
$\sigma = 30 \kms$. Examples of fits are shown in Fig.~\ref{f:transect}.

The distribution of $Q_{\rm max}$ and $N_{\rm max}$ over the GMOS
field-of-view suggests that most fits within the region of the source
where the signal level is high are reliable but some points near the
nucleus have lower reliability, perhaps because the required number of
components exceeds $N_{\rm max}$. This is not surprising in regions
where the signal/noise ratio is very high.  There is a clear trend to
greater numbers of components in the brighter (nuclear) regions of the
object. This is likely to be due to the higher signal/noise in the
brighter regions but may also reflect genuinely more complex
kinematics in the nuclear region.  An atlas of fits for each datapoint
in the field of view with $N \geq 1$ is given in electronic form in
the appendix.

\subsection{Steps towards interpretation}

Garcia-Lorenzo \etal (1999) and Emsellem \etal (2005) identified three
distinct kinematical components in their [O{\small III}] and H$\beta$
data based on line width. Our GMOS data obtained at a finer spatial
and spectral resolution paints a considerably more complex picture. At
even finer spatial resolution, HST spectroscopic data (e.g. Cecil et
al. 2002; Groves et al. 2004) obtained with multiple longslit
positions and narrow-band imaging adds further complexity, including
features not seen in our data in which individual clouds can be
identified and classified by their kinematics.

However, the HST longslit data must be interpreted with caution since
it is not homogeneous in three dimensions and cannot be assembled
reliably into a datacube. Making direct links between features seen in
our data with those of these other authors is not simple, but
subjective and ambiguous. This clearly illustrates how our
understanding of even this accessible, archetypal galaxy is still
strongly constrained by the available instrumentation despite recent
major advances in integral-field spectroscopy and other 3D
techniques.

The atlas of multicomponent fits to the emission line data provides a
huge dataset for which many interpretations are possible.  To indicate
the complexity of the data, the line components along representative
transects through the data are plotted in Figs.~\ref{f:transect2} \&
\ref{f:transect4}. The plots give the radial velocity for each fitted
component as a function of distance along the transect as indicated on
the white-light image obtained by summing over all wavelengths in the
datacube. The component flux (the product of fitted FWHM and
amplitude) is represented by the area of the plotted symbol and the
FWHM is represented by its colour.  H$\beta$ absorption is negligible
compared to the emission and has been ignored in the fits.

Fig.~\ref{f:transect2} is for a ``dog-leg'' transect designed to
encompass the major axis of the structure surrounding the brightest
point in the continuum map (assumed to be the nucleus) and the bright
region to the north-east. For comparison, Fig.~\ref{f:transect4} shows a
transect along an axis roughly perpendicular to this. The following
general points may be made from this:

\begin{figure*}
\includegraphics[height=12.5cm,angle=-90]{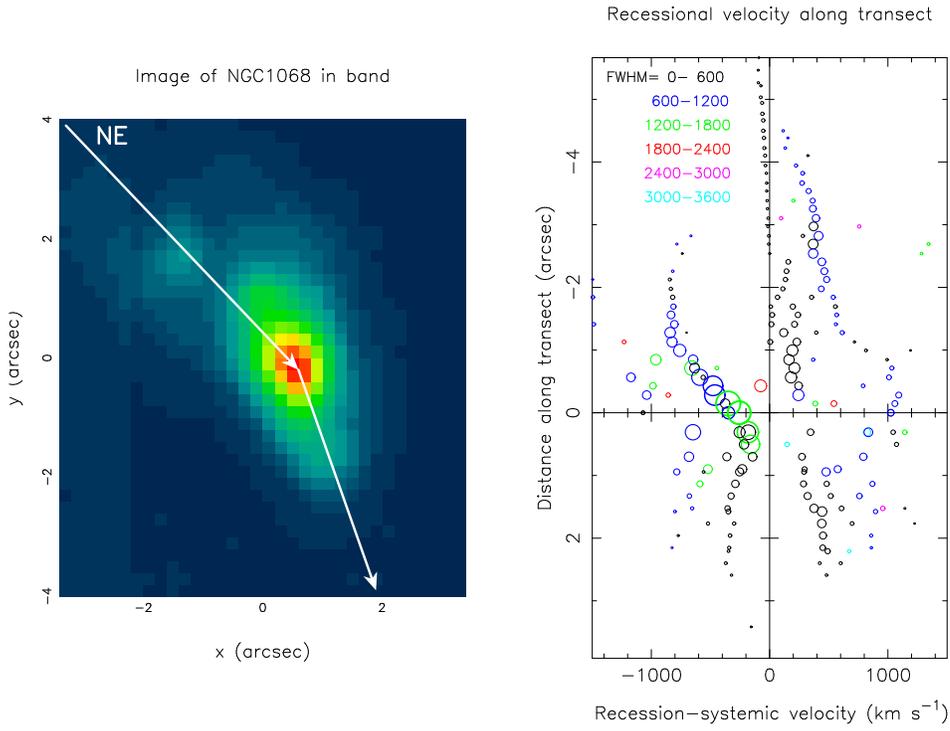}
\caption{A plot of radial velocity versus distance along the dog-leg
NE-SW transect, as shown in the image, for each component fitted to
the H$\beta$ line excluding uncertain fits.  The change of direction
marks the zero-point of the distance scale. The area of the circles is
proportional to the component flux while the colour encodes the line
width as indicated in the key. See the text for further details.}
\label{f:transect2}
\end{figure*}

\begin{figure*}
\includegraphics[height=12.5cm,angle=-90]{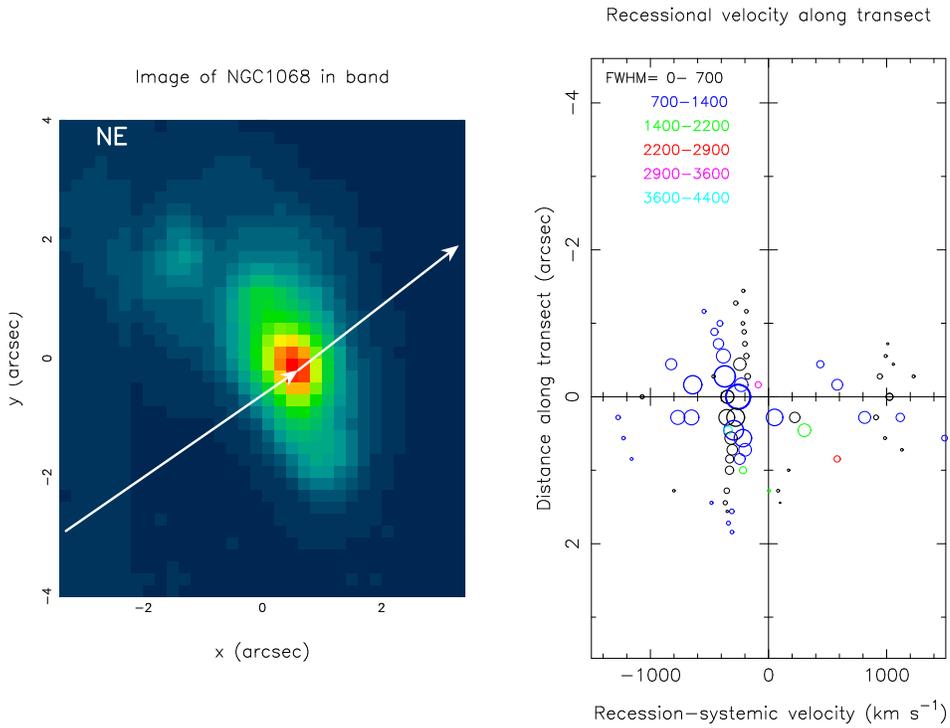}
\caption{Same as the preceding figure but for a transect perpendicular
to the previous one.}
\label{f:transect4}
\end{figure*}

\begin{itemize}

\item Strong shears in radial velocity are present between north-east
and south-west

\item At each position along the transects there are a number of
distinct components with different bulk radial velocity, spanning a
large range ($\sim$3000 km/s)

\item The majority of the observed flux comes from a component spanned
by relative radial velocity --1000~km/s to 0~km/s

\item This component, if interpreted as a rotation curve has a
terminal velocity of $\pm$500 km/s, too large for disk rotation, so is
more likely to indicate a biconical outflow with a bulk velocity that
increases with distance from the nucleus, reaching an asymptotic
absolute value of 500~km/s. This component also has a large dispersion
which is greatest close to the nucleus.

\item The closeness in radial velocity between components of modest
dispersion within this structure (at transect distances between 0 and
+1~arcsec) may arise from uncertainties in the decomposition of the
line profiles into multiple components, i.e. two components of
moderate dispersion are similar to one component with high
dispersion,

\item The other components are likely to indicate rotation of gas
within disk-like structures, if the implied terminal velocities are
small enough, or outflows or inflows.

\item Any such flows do not appear to be symmetric about
either the systemic radial velocity or the bulk of the emission
close to the nucleus.

\end{itemize}

A comparison of the STIS [O{\small III}] emission spectra along
various slit positions (Cecil et al.; Groves et al.) provides
qualitative agreement with our data in that most of the components in
our datacube have analogues in their data. However their data suggest
a much more clumpy distribution than ours, which is not surprising
given the difference in spatial resolution. The apparently much
greater range of radial velocity in their data is because our
component maps indicate the {\it centroid} of the radial velocity of
each component, not the full width of the line, which can be
considerable (FWHM of 1000-2000 km/s). If we plot the line components
in the same way as them, the dominant -1000 to 0 km/s component would
fill in the range --2500 to +1000 in good agreement with their data.

There is evidence of a bulk biconic flow indicative of a jet directed
in opposite directions from the nucleus plus a number of flows towards
and away from the observer which could be either inflowing or
outflowing, but which show a preference to be moving away from the
observer. Some of these could be associated with disk-like components.
As a first attempt towards interpretation, we seek to identify any
gaseous components which can be associated with the stellar
kinematics.

\section{Absorption Line Data}
\label{s:absdat}

\subsection{Stellar Kinematics}

\begin{figure*}
\includegraphics[width=17cm]{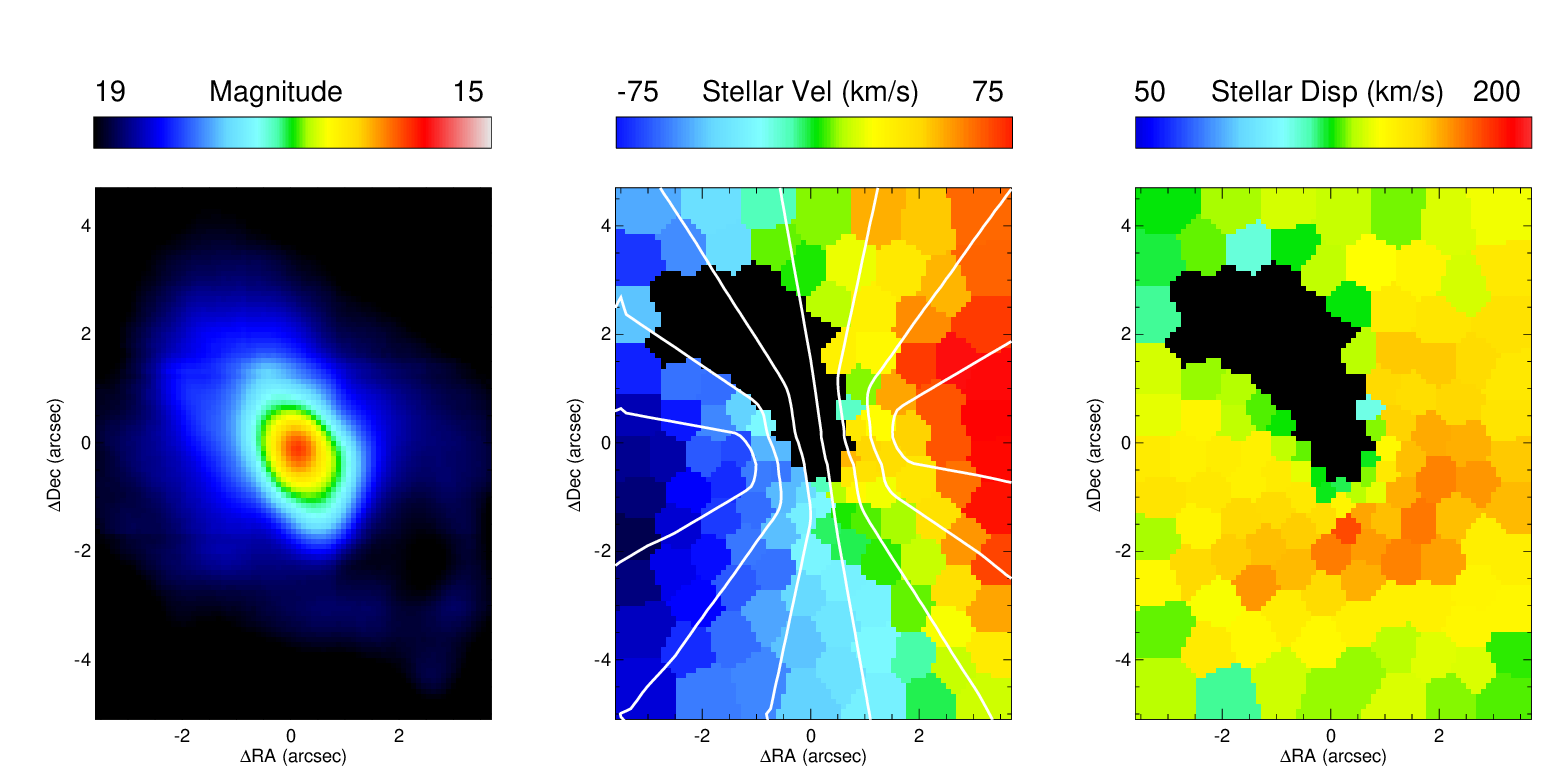}
\caption{Maps of the stellar kinematics in the centre of NGC~1068.
  The reconstructed image (left panel) is obtained by summing all spectra in
  the GMOS IFU datacube from 5100 to 5400 \AA. The magnitude scale has an
  arbitrary offset. The kinematical analysis presented in the next two panels
  uses data that are coadded into spatial bins with a signal-to-noise ratio of
  20 or more using a Voronoi binning algorithm (Cappellari \& Copin, 2003).
  The stellar velocity field (centre) and the stellar velocity dispersion
  (right panel) were derived by fitting a stellar template spectrum (type
  K0III) convolved with a Gaussian distribution to the spectral region around
  the Mg\,$b$ triplet ($\sim 5170$ \AA).  The systemic velocity has been
  subtracted in the central panel.  The white contours in this panel show the
  best-fit model velocity field, see section~\ref{s:diskmodel}. In all panels
  North is to the top and East is to the left.  Blank regions indicate those
  areas where strong non-continuum emission prevented satisfactory fits to the
  absorption line data. } 
\label{f:maps}
\end{figure*}

\begin{figure}
\includegraphics[width=8cm]{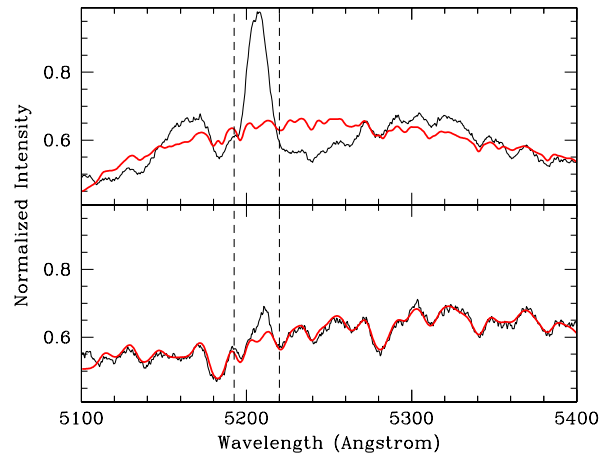}
\caption{Two examples of stellar template fits (bold red lines) to
absorption lines in the data cube. The top panel shows a spectrum from a
location in the `wedge' where strong non thermal emission significantly
`eroded' the absorption lines while the bottom panel shows a typical
absorption line spectrum. In both panels the continuum is fit with a third
order polynomial.  A 6th order polynomial in the top panel would fit the
continuum wiggles better but the derived kinematics are qualitatively
similar to the 3rd order fit.  The emission line in these panels is the
[N{\small I}] line ($\lambda\lambda \ 5199.8$\AA).  The wavelength between
the two dashed lines were excluded in the stellar template fits.  }
\label{f:absfit}
\end{figure}

\begin{figure}
\includegraphics[width=8cm]{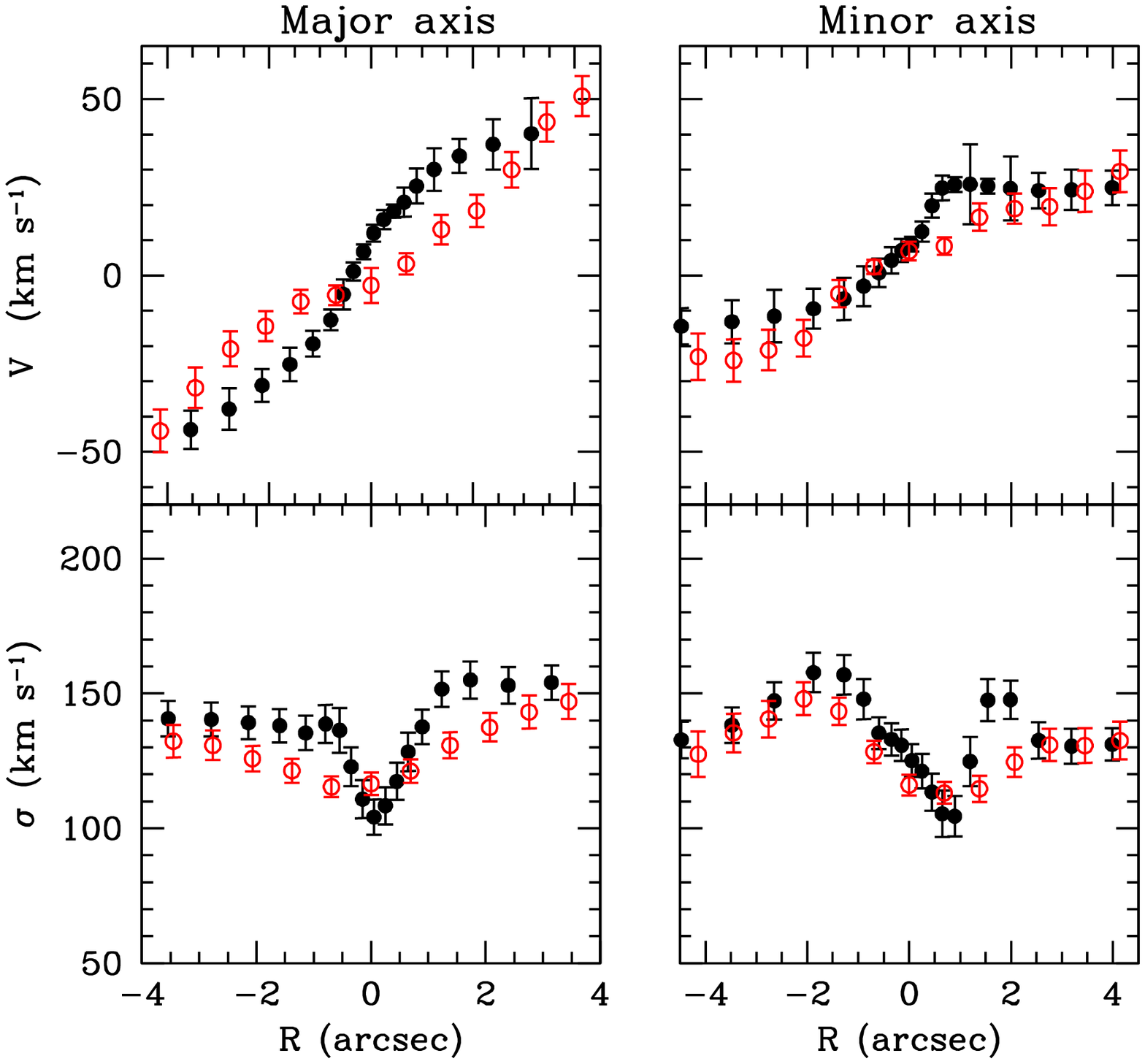}
\caption{A comparison between the results derived from long-slit
observations of Shapiro \etal (2003, open circles) and stellar
kinematics derived from our GMOS data (filled circles). The long-slit
results are obtained with the KPNO 4m telescope along the major axis
(PA$=80^\circ$) and the minor axis (PA$= 170^\circ$) of NGC~1068 (see
Fig.~\ref{f:fov}). The slit width of 3.0 arcsec covers a substantial
fraction of the GMOS observations.  We mimic these long-slit
observations by deriving the GMOS results in 3.0 arcsec wide cuts
through the GMOS data. The differences between the two data sets
are consistent with the effect of different PSFs.}
\label{f:velcomp}
\end{figure}

Although the GMOS IFU commissioning observations of NGC~1068
concentrated on the bright emission lines, the Mg\,$b$ absorption line
feature ($\sim 5170$ \AA) is readily identifiable in the spectra over
most of the field covered by these observations and can be used to
derive the stellar kinematics. In order to minimize contamination by
emission lines when extracting the stellar kinematics, we only used
data in the wavelength range from 5150 to 5400 \AA.

In order to reliably extract the stellar velocities from the
absorption lines it is necessary to increase the signal-to-noise
ratios by co-adding spectra.  The $103 \times 79$ individual spectra
in the data cube were co-added using a Voronoi based binning algorithm
(Cappellari \& Copin 2003).  While the improvement in signal/noise is
at the expense of spatial resolution, the chosen algorithm is
optimised to preserve the spatial structure of the target.  A
signal-to-noise ratio of $\gtrsim 20$ was used to bin the data, yielding 160
absorption line spectra above this threshold.

The stellar kinematics are extracted from these spectra using the
standard stellar template fitting method.  This method assumes that
the observed galaxy spectra can be approximated by the spectrum of a
typical star (the stellar template) convolved with a Gaussian profile
that represents the velocity distribution along the line-of-sight. The
Gaussian parameters that yield the smallest difference in $\chi^2$
between an observed galaxy spectrum and the convolved template
spectrum are taken as an estimate of the stellar line-of-sight
velocity and the line-of-sight stellar velocity dispersion.  Various
codes have been developed over the past decade that implement stellar
template fitting.  We have used the pixel fitting method of
van~der~Marel (1994) in this paper.

Stellar template spectra were not obtained as part of the NGC~1068
GMOS IFU observations. Instead, we used a long-slit spectrum of the
K0III star HD 55184 that was obtained with the KPNO 4m telescope in
January 2002 on a run that measured the stellar kinematics along the
major and minor axis of NGC~1068 (Shapiro \etal 2003).  The
instrumental resolution of the KPNO spectrum is the same as our GMOS
data.  Both the template spectrum and the GMOS IFU spectra were
resampled to the same logarithmic wavelength scale before applying the
kinematical analysis. The continuum in the galaxy spectra is handled
by including a 3rd order polynomial in the fitting procedure.

The best-fit stellar velocities and stellar velocity dispersions are
presented in Fig.~\ref{f:maps}.  A section of the kinematical maps
directly North East of the centre is empty. No acceptable stellar
template fits could be obtained in this region (see also
Fig.~\ref{f:absfit}).  The blank region lies along the direction of the
approaching NE jet. Boosted non-continuum emission is therefore the
most likely explanation for the observed blanketing of the absorption
line features in this area.

\subsection{Comparison and interpretation}

Although part of the stellar velocity map is missing, the overall
appearance is that of a regularly rotating stellar velocity field.
The kinematical minor axis (with radial velocity, $v=0$) is aligned
with the direction of the radio jets and with the long axis in the
reconstructed GMOS IFU image.  Garcia-Lorenzo \etal (1997) derive the
stellar velocity field over the central $24 \times 20$~arcsec in
NGC~1068 from the Ca II triplet absorption lines.  The part of their
velocity map that corresponds to the GMOS IFU data is qualitatively
consistent with our velocity map although there appears to be a
rotational offset of about 15 degrees between the two velocity maps.
Their best-fit kinematical major axis has PA$ = 88^\circ \pm 5^\circ$
while the GMOS IFU maps suggests PA $\simeq 105^\circ$. The latter is
more consistent with the Schinnerer \etal (2000) best-fit major axis
at PA$=100^\circ$.

In Fig.~\ref{f:velcomp} we compare the GMOS data with results we
derived from the long-slit data obtained along the (photometric) major
and minor axes of NGC~1068 (Shapiro \etal 2003).  The long-slit
results were obtained from a 3.0 arcsec wide slit.  The GMOS
kinematics shown in this figure were therefore derived by mimicking
the long-slit observations.  That is, we extracted the kinematics from
3.0 arcsec wide cuts through the GMOS data along the same PAs as the
long-slits.  The long-slits partially overlap the wedge region seen in
Fig.~\ref{f:maps}. However, because of the width of the mimicked
slits, the contaminating effect of the centrally concentrated emission
lines was significantly reduced.  But some effects on the derived
stellar kinematics remain.  Most notably along the minor axis profiles
at $\sim 2$ arcsec.  (The Shapiro \etal results were obtained in
fairly poor seeing conditions resulting in additional smearing.)

The stellar velocity dispersion map is missing the same region as the
stellar velocity map. Although this includes the nucleus, it appears
that the maximum velocity dispersion is located off-centre. The
velocity dispersion profiles of NGC~1068 published in Shapiro \etal
(2003) focused on the behaviour at large radii where contamination by
emission lines is not an issue.  A re-analysis of these data using a
pixel based method rather than a Fourier based method shows that the
velocity dispersions in NGC~1068 exhibit the same central drop (see
also Emsellem \etal 2005) observed in the GMOS data.  As the spectra
of NGC~1068 show very strong emission lines in the central region of
this system, the difference with the Shapiro \etal results can be
attributed to unreliability of Fourier-based methods in the presence
of significant emission lines.

Assuming that the velocity dispersions are distributed symmetrically
around the major axis (PA$=100^\circ$) of the kinematical maps, the
dispersion distribution resembles the dumbbell structures found in the
velocity dispersion maps of the SABa type galaxy NGC~3623 (de Zeeuw
\etal 2002) and the SB0 type galaxies NGC~3384 and NGC~4526 (Emsellem
\etal 2004). A rotating (i.e.  cold) disk embedded in a bulge
naturally produces the observed dumbbell structure in the velocity
dispersions.  Both the position angle of the kinematical major axis
and the orientation of the inferred nuclear disk are consistent with
the central CO ring (diameter: 5~arcsec) identified by Schinnerer
\etal (2000). Alternative interpretation of the incomplete NGC~1068
stellar dispersion map include a kinematically decoupled core or
recent star formation (e.g. Emsellem \etal 2001).  The GMOS stellar
velocity maps compliment the larger scale maps of Emsellem \etal
(2005) where the central structure (see their Fig.~6) becomes
uncertain for the same reason as for the empty portion of our data.

\subsection{A stellar disk model}
\label{s:diskmodel}

Both the gas and stellar data present a very rich and complex
morphology.  As a first step toward interpretation we fit the GMOS
stellar velocity map with a qualitative `toy' model of a rotating
disk.  In a forthcoming paper, much more detailed and quantitative
models of the gas and stellar data will be presented.

The disk model consists of an infinitely thin circular disk (in this simple
model we use a constant density profile) with a rotation curve that rises
linearly to $75 \kms$ out a radius of 3 arcsec and thereafter remains
constant.  Both the best-fit amplitude and the break radius as well as the
position angle of the disk model were found empirically by comparing the model
velocity field to the GMOS velocity map in a least-square sense.  We assumed
that the disk is parallel to the plane of the galaxy (i.e. we assumed an
inclination of $i=30$ degrees).  The model velocity field is overplotted with
white contours in the central panel of Fig.~\ref{f:maps}.  The best-fit
position angle of the disk model differs by some 30$^\circ$ (PA$_{\rm disk} =
110^\circ$) from the major axis of NGC~1068 but is in close agreement with the
kinematical position angle (e.g. Schinnerer \etal 2000).

\begin{figure*}
\includegraphics[width=14cm]{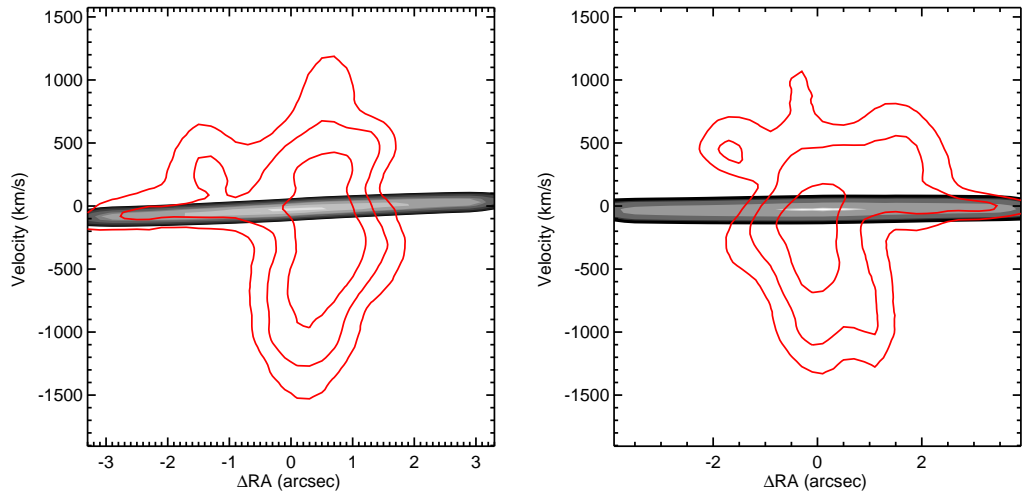}
\caption{Comparison between the projected H$\beta$ flux distribution
(thin lines) observed in the centre of NGC~1068 with the GMOS IFU and
the projected density distribution of the disk model (thick shaded
lines). In the left panel the data cube is collapsed along the
declination axis, while the panel on the right shows the data cube
collapsed along the RA axis. The disk-like structure in these data is
clearly visible as the spatially extended, but narrow in velocity
space, structure.  The disk model closely matches both the location
and the gradient of this structure. }
\label{f:modcomp}
\end{figure*}

A comparison between the disk model and the H$\beta$ data is shown in
Fig.~\ref{f:modcomp} at two different projections.  The thin contours
show the projected intensity distribution of H$\beta$ as a function of
position and velocity.  The projected intensity distribution of the
disk model is overplotted as thick lines.  In both panels the H$\beta$
distribution shows a spatially extended structure that is narrow in
velocity space.  The overplotted disk model closely matches the
location and gradient of this structure in the data cube.  To facilitate
a comparison by eye the velocity distribution of the rotating disk
model was broadened by $140 \kms$, the mean observed velocity
dispersion, in this figure.  The most straightforward interpretation
of this result is that part of the emission line distribution
originates in a rotating gas disk that is aligned with the stellar
disk in the centre of NGC~1068.  This component can, with hindsight,
clearly be seen in the transect maps (most evident in
Fig.~\ref{f:transect2} near transect position $-4$ arcsec) as the
aligned set of narrow components that are located closely to the
systemic velocity of NGC~1068.

\section{Conclusions}

We have used integral field spectroscopy to study the structure of the
nucleus of NGC1068 at visible wavelengths. We present an atlas of
multicomponent fits to the emission lines. Interpretation of this
large, high-quality dataset is not straightforward, however, since the
link between a given fitted line component and a real physical
structure cannot be made with confidence. We have started our
exploration of the structure by using absorption line data on the
stellar kinematics in the disk to identify a similar component in the
emission line data which presumably has its origin in gas associated
with the disk. This analysis serves to illustrate both the complexity
of the source and the enormous potential of integral-field
spectroscopic data to understand it as well as the need for better
visualization and analysis tools.

\section*{Acknowledgments}

We thank the anonymous referee for the very detailed comments and
suggestions that helped improve the paper and the SAURON team for the
use of their colour table in Fig.6.  We acknowledge the support of the
EU via its Framework programme through the Euro3D research training
network HPRN-CT-2002-00305.  The Gemini Observatory is operated by the
Association of Universities for Research in Astronomy, Inc., under a
cooperative agreement with the NSF on behalf of the Gemini
partnership: the National Science Foundation (United States), the
Particle Physics and Astronomy Research Council (United Kingdom), the
National Research Council (Canada), CONICYT (Chile), the Australian
Research Council (Australia), CNPq (Brazil) and CONICET (Argentina).

\label{lastpage}

\end{document}